\documentclass[aps,prb,twocolumn,superscriptaddress,nofootinbib]{revtex4-2}
\usepackage[T1]{fontenc} 
\usepackage{float}
\usepackage{graphicx}
\usepackage{dcolumn}
\usepackage{bm}
\usepackage{color}
\usepackage{amsmath}
\usepackage{amssymb}
\usepackage{dsfont}
\usepackage{xcolor}
\usepackage{mathrsfs}
\usepackage{todonotes}
\usepackage{braket}
\usepackage[colorlinks = true,
            linkcolor = blue,
            urlcolor  = red,
            citecolor = blue,
            anchorcolor = blue]{hyperref}
\usepackage{ulem}
\usepackage{soul}

\usepackage{mathdots}
\newcommand{\pdag}{{\phantom{\dagger}}}
\newcommand{\eph}{{\it e}-ph}

\newcommand{\D}{\mathrm{d}}
\newcommand{\Tc}{T_{\text{c}} }

\newcommand{\cc}{\mathrm{c}}

\newcommand{\kB}{k_{\mathrm{B}}}
\newcommand{\ra}{\rangle}
\newcommand{\la}{\langle}
\newcommand{\ua}{\uparrow}
\newcommand{\da}{\downarrow}

\newcommand{\e}{\mathrm{e}}
\newcommand{\ii}{\mathrm{i}}
\newcommand{\bk}{\mathbf{k}}

\newcommand{\bq}{\mathbf{q}}

\newcommand{\br}{\mathbf{r}}

\newcommand{\bx}{\mathbf{x}} 
\newcommand{\by}{\mathbf{y}}

\newcommand{\qp}{q^{\prime}}
\newcommand{\tp}{t^{\prime}}
\newcommand{\mpr}{m^{\prime}}

\newcommand{\bqp}{\mathbf{q}^{\prime}}

\newcommand{\cdag}{\hat{c}^{\dagger}}
\newcommand{\chat}{\hat{c}^{\pdag}}

\newcommand{\Deldag}{\hat{\Delta}^{\dag}}
\newcommand{\Delhat}{\hat{\Delta}^{\pdag}}

\newcommand{\sigp}{\sigma^{\prime}}

\newcommand{\mrm}[1]{\mathrm{#1}}

\newcommand{\Ttau}{ \hat{ \mathscr{T} }_{\tau} }	
\allowdisplaybreaks

\newcommand{\deltaFD}{ \tilde{\delta}_{\mrm{fd}} }

\begin{document}

\title{Charge correlations suppress unconventional pairing in the Holstein model}
\author{Philip M. Dee}
\affiliation{Department of Physics, University of Florida, Gainesville, Florida 32611, USA}
\affiliation{Department of Materials Science and Engineering, University of Florida, Gainesville, Florida 32611, USA\looseness=-1}
\author{Benjamin Cohen-Stead}
\affiliation{Department of Physics and Astronomy, The University of Tennessee, Knoxville, Tennessee 37996, USA}
\affiliation{Institute of Advanced Materials and Manufacturing, The University of Tennessee, Knoxville, Tennessee 37996, USA\looseness=-1}
\author{Steven Johnston}
\affiliation{Department of Physics and Astronomy, The University of Tennessee, Knoxville, Tennessee 37996, USA}
\affiliation{Institute of Advanced Materials and Manufacturing, The University of Tennessee, Knoxville, Tennessee 37996, USA\looseness=-1}
\author{P. J. Hirschfeld}
\affiliation{Department of Physics, University of Florida, Gainesville, Florida 32611, USA}
\date{\today}

\begin{abstract}
In a recent work by Schrodi \textit{et al.} [Phys. Rev. B. \textbf{104}, L140506 (2021)], the authors find an unconventional superconducting state with a sign-changing order parameter using the Migdal-Eliashberg theory, including the first vertex correction. This unconventional solution arises despite using an isotropic bare electron-phonon coupling in the Hamiltonian. We examine this claim using hybrid quantum Monte Carlo for a single-band Holstein model with a cuprate-like noninteracting band structure and identical parameters to Schrodi \textit{et al.}. Our Monte Carlo results for these parameters suggest that unconventional pairing correlations do not exceed their noninteracting values at any carrier concentration we have checked. Instead, strong charge-density-wave correlations persist at the lowest accessible temperatures for dilute and nearly half-filled bands. Lastly, we present arguments for how vertex-corrected Migdal-Eliashberg calculation schemes can lead to uncontrolled results in the presence of Fermi surface nesting.
\end{abstract}

\maketitle
\section{Introduction}
%
The possible role of electron-phonon ($e$-ph) interactions in high-temperature (high-$\Tc$) superconductors is a long-standing problem. 
Coupling at small momentum transfer, $\bq$, can lead to attractive interactions in unconventional pairing channels~\cite{Mott1993, KulicPRB1994, Santi1996, VarelogiannisPRB1996, DevereauxPRL2004, Kulic2005, JohnstonPRL2012, LeeNature2014, WanNatCommun2014, RademakerNJP2016}. 
There are also theoretical studies suggesting that the $e$-ph coupling can be enhanced at small $\bq$ transfers by the Coulomb interaction through screening \cite{HuangPRB2003, JohnstonPRB2010} and anisotropy in the transport properties \cite{JohnstonPRB2010, LeeNature2014}. 
In these scenarios, the momentum structure of the $e$-ph coupling constant $g({\bf k},{\bf q})$ gives rise to attractive contributions  $\lambda_l$ in multiple angular momentum channels. 
For any realistic $g({\bf k},{\bf q})$, the coupling in the $s$-wave channel is dominant, and the interaction will lead to an $s$-wave order parameter in the absence of any repulsive interactions. 
However, strong repulsive interactions like a large Hubbard $U$ or $\mu^*$ can suppress $s$-wave pairing in favor of an unconventional pairing symmetry \cite{ScalapinoRMP, HirschfeldReview}.
Once this occurs, the next leading order contribution from the $e$-ph interaction can provide an additional boost to the pairing glue, provided it is an attractive interaction in the appropriate pairing channel (e.g., $\lambda_{x^2-y^2}$ for cuprates or $\lambda_{\pm s}$ for the Fe-based superconductors).

\begin{figure}[t]
	\includegraphics[width=1.0\columnwidth]{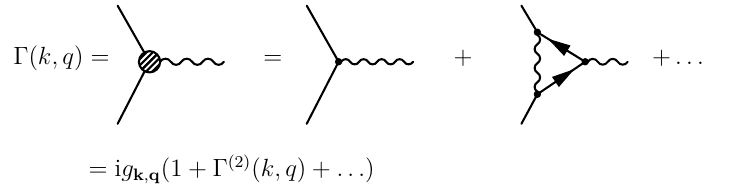}
	\caption{The electron-phonon vertex of a simple electron-phonon system 
	$\Gamma(k,q)\equiv \Gamma(\bk,\ii\omega_{n};\bq,\ii\nu_{m})$ is a sum of Feynman diagrams where the first term is simply the bare vertex $\ii g_{\bk,\bq}$. 
	The ``first vertex correction'' is given by $\Gamma^{(2)}(k,q)$. Higher order 
	diagrams are not considered in this work. 
	}
	\label{fig:vertex_diagrams}
\end{figure}

Recently, Schrodi~\textit{et al.}~\cite{SchrodiPRB2021} proposed that a Holstein interaction\textemdash i.e., a momentum independent $e$-ph interaction\textemdash can mediate an attractive interaction in unconventional channels \textit{without} the additional influence of electron correlations. 
Those authors examined several models, including a single-band Holstein model for the high-$\Tc$ cuprates, as well as multiband models for the Fe-based, and heavy-fermion superconductors with nested Fermi surfaces. 
In each case, they considered a Holstein $e$-ph coupling within a vertex-corrected Eliashberg-theory calculation (see Fig.~\ref{fig:vertex_diagrams}), where the rainbow and first vertex correction diagrams for the electron self-energy are computed self-consistently. 
In doing so, they found that the inclusion of the vertex corrections leads to instabilities in \textit{unconventional} pairing channels. 
Moreover, the symmetry of the derived order parameter in each case was consistent with those derived from weak coupling repulsive spin-fluctuation-based models and Fermi surface nesting arguments~\cite{MazinReview, ScalapinoRMP}.  

The results of Schrodi~\textit{et al.}~\cite{SchrodiPRB2021} are at odds with many nonperturbative studies of the single-band Holstein model, which find that the temperature-doping phase diagram is dominated by charge-density-wave or $s$-wave pairing correlations~\cite{FreericksPRB1993, ScalettarPRB1989, MeyerPRL2002, CaponePRL2003, CaponePRB2006, TezukaPRB2007, MurakamiPRL2014, KarakuzuPRB2017, OhgoePRL2017, ChenPRB2018, HohenadlerPRB2019, LiPRB2020, EsterlisPRB2018, Dee2020, BradleyPRB2021, NosarzewskiPRB2021, PaleariPRB2021}. 
Here, we explicitly explore their claim using a state-of-the-art hybrid Monte Carlo (HMC) method~\cite{CohenSteadPRE2022}. 
Specifically, we obtain numerically exact solutions to the same cuprate model examined in Ref.~\citenum{SchrodiPRB2021}.  
The model is dominated by charge-density-wave (CDW) correlations down to the lowest temperatures we examine, which overlap the range studied by Ref.~\citenum{SchrodiPRB2021}.
Performing simulations at fixed values of the electronic density reveals that the pairing correlations
are largely suppressed below their noninteracting counterparts, regardless of the leading pairing
symmetry.” 
Alternatively, when simulations are performed for a fixed chemical potential, we find the bands shift above the Fermi level as the temperature decreases, indicating that the self-energy effects from the $e$-ph coupling are substantial. 
At no point do we observe an instability toward a superconducting phase with an unconventional order parameter. 
With this result in mind, we then examine the momentum structure of the first vertex correction and argue that truncating the expansion at the first vertex correction is an uncontrolled approximation when the Fermi surface is well nested. 



%
%
\section{Model and Methods}
We study a single-band Holstein model, defined on a two-dimensional square lattice. 
The Hamiltonian is given by
\begin{align}\label{eq:HHolstein}\nonumber
    \hat{H} =&-\sum_{i,j} t_{i,j}^\pdag \cdag_{i,\sigma} \chat_{j,\sigma} - \mu\sum_{i,\sigma} \hat{n}_{i,\sigma} 
    +\sum_{i} \left[\frac{\hat{P}^{2}_i}{2M}+\frac{M\Omega^{2}}{2} \hat{X}_i^2\right] \\
    &+ \alpha\sum_{i,\sigma}  \hat{n}_{i,\sigma}\hat{X}_{i},
\end{align}
where $\cdag_{i,\sigma}$ creates a spin-$\sigma$ ($=\uparrow,\downarrow)$ 
electron on site $i$, $\hat{n}_{i,\sigma} = \cdag_{i,\sigma}\chat_{i,\sigma}$ 
is the Fermion number operator for site $i$, $t_{i,j}$ is the hopping integral 
between sites $i$ and $j$, $\mu$ is the chemical potential, $\hat{X}_i$ and $\hat{P}_i$ are 
the position and momentum operators for the atom at site $i$, $M$ is the ion 
mass, $\Omega$ is oscillator frequency, and $\alpha$ is the 
\eph~coupling strength. 
The single-band tight-binding dispersion $\epsilon_{\bk}$ for this model is given by 
$\xi_\bk = \epsilon_{\bf k}-\mu$, where 
\begin{equation}
    \epsilon_{\bk} = -2t [\cos(k_{x}) + \cos(k_{y})] 
    - 4\tp [\cos(k_{x}) \cos(k_{y}) ]
\end{equation}
and we have set the lattice spacing $a=1$.

Throughout, we set $M = \hbar = 1$ such that the energy of the phonon modes 
$\hbar\Omega\rightarrow\Omega$, and restrict the hopping to nearest- ($t$) and 
next-nearest-neighbors ($t^\prime$), only. 
We then adopted $t = 1$, $t^\prime/t = -0.2$, $\Omega/t = 0.4$, and $\alpha = 1.059$, following Ref.~\citenum{SchrodiPRB2021}.\footnote{Ref.~\citenum{SchrodiPRB2021} defines the bare band structure as $\xi_\bk = -t\left[\cos(k_xa)+\cos(k_ya)\right]-t^\prime \cos(k_xa)\cos(k_ya)-\mu$. We have, therefore, selected our $t$ and $t^\prime$ values to match their bare band dispersion.}
These values result in a large dimensionless \eph~coupling of 
$\lambda = \alpha^{2}/(W\Omega^{2}) \approx  0.88$, where  $W \approx 8t$ is the noninteracting bandwidth.
The chemical potential $\mu$ controls the filling in our simulations. 
Later, we will show results for fixed $\mu/t = -0.56$ and as well as for a fixed average filling $n\equiv\langle \hat{n} \rangle = \frac{1}{N}\sum_{i,\sigma}\langle \hat{n}_{i,\sigma}\rangle = 0.8$ and 0.2.
In the latter cases, $\mu$ is determined dynamically within the HMC simulation using a recently developed $\mu$-tuning algorithm~\cite{MilesPRE2022}. 

We solve the model using a recently developed method~\cite{CohenSteadPRE2022}, which leverages HMC~\cite{Duane1987, BeylPRB2018} and Fourier acceleration to reduce decorrelation time of the phonon fields~\cite{BatrouniPRB2019}, a physics-inspired preconditioner, and near-linear scaling measurement techniques. 
This approach allows us to simulate large system sizes and consider optical phonons with energies much smaller than the electron hopping and equal to those used in Ref.~\citenum{SchrodiPRB2021}. 
We performed all of our HMC simulations on $N = 12\times 12$ clusters.

The strength of the charge correlations is determined by measuring the charge structure factor
\begin{equation}\label{eq:structure_factor}
    S(\bq,\tau) = \frac{1}{N} \sum_{i,j} \e^{-\ii\bq\cdot(\br_{i} - \br_{j})} 
    \la \Ttau[\hat{n}_{i}(\tau)\hat{n}_{j}(0)] \ra,
\end{equation}
where $\Ttau$ is the time-ordering operator, and charge susceptibility
\begin{equation}\label{Eq:chi_cdw}
    \chi^\mathrm{CDW}(\bq) = \int_0^\beta S(\bq, \tau) \ \D\tau. 
\end{equation}
The strength of the pairing correlations is determined by the pair-field susceptibility
\begin{equation}\label{eq:chi_pair}
  \chi_{\alpha}^{\mrm{SC}} = \frac{1}{N} \int_{0}^{\beta} \D\tau\,
 	\la 
 		\Ttau[\Delhat_{\alpha}(\tau) \Deldag_{\alpha}(0)]
 	\ra, 
\end{equation}
for some pairing symmetry 
$\alpha=s,\,d,\,p$, etc., the operator $\Deldag_{\alpha}$ is defined~\cite{Nowadnick2015} as 
\begin{equation}\label{eqn:Deldag_alpha(tau)}
 	\Deldag_{\alpha} = \frac{1}{2}\sum_{i,\gamma}f_{\gamma}^{\alpha}
 		\cdag_{i,\ua} \cdag_{i+\gamma,\da}.
\end{equation}
Here, the sum over $\gamma$ is restricted up to nearest neighbors only. 
For $s$-wave pairing, $f_\gamma^{s}=\delta_{\gamma,0}$, where $\delta_{i,j}$ is the usual Kronecker-delta. 
For $d$-wave pairing, $f_{\gamma}^{d} =  \delta_{\gamma,\pm\hat{\bx}}-\delta_{\gamma,\pm\hat{\by}}$.
%

\section{Results}
\subsection{Hybrid Monte Carlo}

\begin{figure}[t]
	\includegraphics[width=1.0\columnwidth]{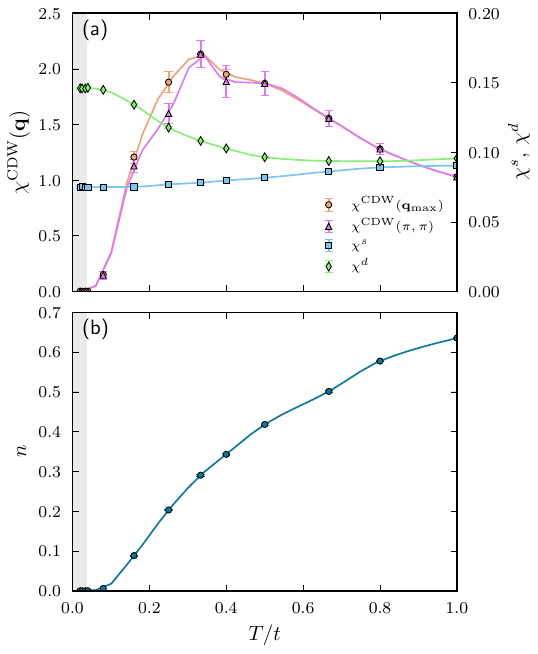}
	\caption{
        The temperature evolution of the (a) charge and superconducting pair-field susceptibilities and 
        (b) electronic filling $n\equiv\la\hat{n}\ra$ in the Holstein model for a fixed chemical potential $\mu/t=0.56$. 
        The largest filling at $T/t = 2$ is $ n  = 0.809$ (outside the plot window). 
        As the temperature is lowered, the band shifts to energies above the Fermi 
        level, and the band is depleted $ n  \rightarrow 0$. Results were obtained on $N = 12\times 12$ clusters.
        The gray shaded region below $T/t = 0.036$ indicates the first vertex-corrected superconducting phase, as predicted by Ref.~\cite{SchrodiPRB2021}.
        $\chi^\mathrm{CDW}(\bq_{\mrm{max}})$ corresponds to the maximum value $\chi^{\mrm{CDW}}$ located at 
        some $\bq=\bq_{\max}$ which can be different from $\bq=(\pi,\pi)$.
        } \label{fig:Susceptibilities_vs_T_fixed_mu}
\end{figure}


Reference~\cite{SchrodiPRB2021} is unclear in how it treats the filling of the system, specifically whether $\mu$ or $n$ is held fixed during the self-consistency loop of their calculations. 
We will consider both cases in what follows. 

We begin with a fixed chemical potential, which we set to $\mu/t = -0.56$ as indicated by Ref.~\cite{SchrodiPRB2021}. 
Figure~\ref{fig:Susceptibilities_vs_T_fixed_mu}(a) plots the evolution of the charge $\chi^\mathrm{CDW}({\bf q})$ and pairing correlations as a function of temperature. 
We find that $\chi^\mathrm{CDW}({\bf q})$ is most prominent at ${\bf q} = (\pi,\pi)$ for nearly all temperatures but displays nonmonotonic behavior taking on a maximum at $T/t\approx 0.4$ before it turns over and rapidly decays to zero. 
At these lowest temperatures, the $d$-wave pair-field susceptibility is indeed larger than the $s$-wave, but neither are significantly larger than their values at high temperature, indicating no strong tendency to pairing. 
The nonmonotonicity in $\chi^\mathrm{CDW}({\bf q})$ occurs because the filling of the system is not fixed and $ n \rightarrow 0$ as the temperature is lowered [Fig.~\ref{fig:Susceptibilities_vs_T_fixed_mu}(b)]. 
This behavior is likely due to significant growth in the self-energy, which shifts the bands above the Fermi level. 
Regardless of the origin, we find no evidence for a $d$-wave instability when we simulate the system with a fixed chemical potential.

\begin{figure}[t]
	\includegraphics[width=1.0\columnwidth]{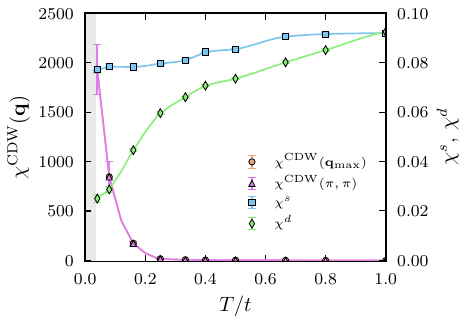}
	\caption{The temperature evolution of the charge and superconducting pair-field susceptibilities in the Holstein model at a       fixed filling $n=0.8$. The remaining parameters are  $t^\prime/t = -0.2$, $\Omega/t = 0.4$, and $\lambda =      0.8762$. Results are obtained on $N = 12\times 12$ clusters.}
	\label{fig:Susceptibilities_vs_T_(B)}
\end{figure}

\begin{figure}[h]
	\includegraphics[width=1.0\columnwidth]{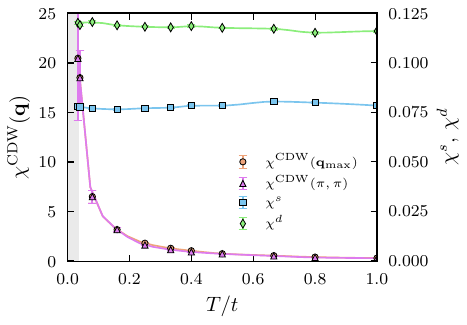}
	\caption{The temperature evolution of the charge and superconducting pair-field susceptibilities in the Holstein model at a fixed filling $n=0.2$. The remaining parameters are 
	$t^\prime/t = -0.2$, $\Omega/t = 0.4$, and $\lambda =  0.8762$. Results are obtained on $N = 12\times 12$ clusters.}
	\label{fig:Susceptibilities_vs_T_(C)}
\end{figure}

\begin{figure}[h]
	\includegraphics[width=1.0\columnwidth]{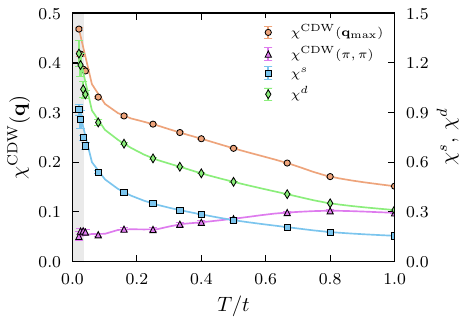}    
	\caption{For reference, we include the temperature evolution of the charge and superconducting pair-field susceptibilities in the \textit{noninteracting} case at a fixed filling $n=0.2$. The remaining parameters are 
	$t^\prime/t = -0.2$, $\Omega/t = 0.4$, and $\lambda =  0$. Results are obtained on $N = 12\times 12$ clusters.}
	\label{fig:NonIntSusceptibilities_vs_T_(C)}
\end{figure}

Next, we fix the average filling to $n = 0.8$, corresponding to the approximate band filling for $\mu/t = -0.56$ obtained at $T/t = 2$. 
Figure~\ref{fig:Susceptibilities_vs_T_(B)} plots the corresponding temperature dependence of the charge and pair-field susceptibilities in this case. 
Here the ${\bf q} = (\pi,\pi)$ charge susceptibility dominates at all temperatures and is up to five orders of magnitude larger than both the $s$- and $d$-wave pair-field susceptibilities for $T/t < 0.2$. 
We can conclude that the low-temperature ground state of the system is dominated by charge correlations and is not superconducting. 
Turning to the superconducting correlations, we find that $\chi^s > \chi^d$ at all temperatures, with the latter dropping significantly once the CDW correlations begin to dominate. 
We find no evidence for enhanced $d$-wave pairing or a superconducting instability for this filling. 
It should also be noted that the vertex-corrected Migdal-Eliashberg calculations of Ref.~\cite{SchrodiPRB2021} placed the superconducting transition at $T_\mathrm{c}/t = 0.036$ (52 K), which falls within our simulation temperatures. 
For easier comparison, we have included a light-gray shaded region at and below their reported $\Tc$ in both panels 
of Fig.~\ref{fig:Susceptibilities_vs_T_fixed_mu} as well as Figs.~\ref{fig:Susceptibilities_vs_T_(B)}-\ref{fig:NonIntSusceptibilities_vs_T_(C)}.

Our results demonstrate that charge correlations are dominant in the Holstein model near half-filling, in agreement with many prior numerical studies~\cite{FreericksPRB1993, ScalettarPRB1989, CaponePRB2006, KarakuzuPRB2017, ChenPRB2018, HohenadlerPRB2019, LiPRB2020, EsterlisPRB2018, BradleyPRB2021, NosarzewskiPRB2021, PaleariPRB2021}. 
Many of those same studies also find strong superconducting correlations for carrier concentrations away from half-filling. 
Motivated by this, we also performed calculations for a dilute filling $n = 0.2$.  
Fig.~\ref{fig:Susceptibilities_vs_T_(C)} plots the resulting temperature evolution of the charge and pairing susceptibilities in this case. 
Here, the noninteracting Fermi surface is free-electron-like (circular) and is far from any nesting conditions.
Nevertheless, we find that ${\bf q} = (\pi,\pi)$ charge correlations dominate the system at low temperatures, while the superconducting correlations remain weak over the temperatures we can access. 
In this case, the large value of $\chi^\mathrm{CDW}(\pi,\pi)$ reflects a strong tendency towards bipolaron formation \cite{EsterlisPRB2018, NosarzewskiPRB2021}, which is not unexpected given the large value of $\lambda$. 
Our results show that the Holstein model for the parameters considered in Ref.~\cite{SchrodiPRB2021} is dominated by bipolaron formation at all carrier concentrations, which tend to order and localize for this value of $\lambda$. 
We find no indications that they condense into a superconducting state of any symmetry. 

In Figs.~\ref{fig:Susceptibilities_vs_T_fixed_mu} and \ref{fig:Susceptibilities_vs_T_(C)}, one can see that although the $\chi^{d}>\chi^{s}$, the values are small overall. 
Moreover, the appearance of a larger $\chi^{d}$ is somewhat misleading.
To demonstrate why, we include noninteracting results for $n=0.2$ down to low temperatures, shown in Fig.~\ref{fig:NonIntSusceptibilities_vs_T_(C)}. 
Compared with Fig.~\ref{fig:Susceptibilities_vs_T_(C)}, the pairing susceptibilities in the noninteracting case are generally larger than the interacting case, and the $d$-wave pairing susceptibility has a higher baseline value than the $s$-wave case.
The expected signature of a superconducting transition is a rapid increase in the pairing susceptibility with decreasing temperature with a magnitude much larger than the noninteracting result.
In such a case, one could attempt a finite-size scaling analysis~\cite{BradleyPRB2021}, but in the absence of such a low-$T$ divergence in the 12$\times$12 case presented here, it does not seem appropriate.

Our results contradict those of Ref.~\cite{SchrodiPRB2021}, which obtained a $d$-wave superconducting solution. 
There are several contributions to this discrepancy, but one particularly important factor is that their calculations do not include the renormalization of the phonon propagator. 
This approximation is, understandably, motivated by a need to reduce the computational complexity stemming from the inclusion of the vertex correction.
However, this approximation is severe, as it prevents CDW correlations driven by conventional phonon softening from growing large enough to compete with superconductivity. 
Schrodi \textit{et al}.~\cite{SchrodiPRB2021_ph_renorm} and others~\cite{MarsiglioPRB1990,EsterlisPRB2018,DeePRB2019} have included these effects in previous calculations without the first vertex correction. 
All have found that including the phonon self-energy in a self-consistent manner reintroduces the tendency toward a charge instability, especially for a nested Fermi surface. 
Our numerically exact solutions include these phonon self-energy effects, which may account for our results' discrepancies. 
However, it is noteworthy that Ref.~\cite{SchrodiPRB2021} also obtained unconventional order parameters using a momentum-independent $e$-ph interaction in two other systems with well-nested Fermi surfaces. 
This observation motivates us to examine the structure of the first vertex correction as a function of nesting in order to assess whether truncating at this order is a controlled approximation. 

\subsection{Analysis of the first vertex correction}
The Feynman diagram for the first-order correction to the bare $e$-ph interaction vertex $g_{\bk,\bq}$ is shown in the rightmost diagram of Fig.~\ref{fig:vertex_diagrams}. 
It is given by 
\begin{multline}\label{eqn:Gamma^(2)(k,q)}
\Gamma^{(2)}(k,q) = \frac{\kB T}{N\hbar^{3}}\sum_{\qp,\sigp}|g_{\bk,\bqp}|^2 
    D_{0}(\qp) G_{0}(k-q-\qp) \\
    \times G_{0}(k-\qp).
\end{multline}
Here, we use the shorthand notation $k\equiv(\bk,\ii\omega_{n})$ and 
$q\equiv (\bq,\ii\nu_{m})$ with fermionic and bosonic Matsubara frequencies 
given by $\omega_{n}=(2n+1)\pi\kB T/\hbar$ and $\nu_{m}=2\pi m\kB T/\hbar$ (with 
$n,m\in \mathds{Z}$), respectively.
Equation~(\ref{eqn:Gamma^(2)(k,q)}) follows directly\footnote{Our specific choice 
of momenta arguments is readily seen by examining the vertex in the context of the 
first self-energy crossing diagram, the latter of which is the second-order correction 
$\Sigma^{(2)}$. The incoming and outgoing fermionic lines are labeled by $k$, the 
first phonon line is labeled by $q$, and the second is labeled by $\qp$. } 
from its Feynman diagram and contains the $e$-ph coupling matrix elements 
$g_{\bk,\bqp}$, the noninteracting phonon propagator
\begin{equation}
    D_{0}(\qp) \equiv D_{0}(\bqp, \ii\nu_{\mpr}) = 
    - \frac{2\Omega_{\bqp}}{\nu_{\mpr}^{2} + \Omega_{\bqp}^{2}},    
\end{equation}
and two noninteracting electron propagators where, for example,   
\begin{align}
    G_{0}(k-\qp) &\equiv G_{0}(\bk-\bqp ,\ii\omega_{n}-\ii\nu_{\mpr}) \nonumber
    \\
    &= \frac{1}{\ii(\omega_{n}-\nu_{\mpr}) - \hbar^{-1}   \xi_{\bk-\bqp}}, 
\end{align}
and $\xi_{\bk}\equiv\epsilon_{\bk} - \mu$. 
We have suppressed spin and band indices since we are working with a 
single-band model with parity in the up and down spin 
directions (e.g., $G_{\ua}=G_{\da}$).
Since we are only interested in comparing the relative strength of the bare 
vertex to the vertex correction, we will work exclusively in units such that 
$\kB=\hbar=M=1$.  
%

For a Holstein model, the phonon dispersion is Einstein-like, and the bare 
$e$-ph coupling is isotropic; hence, $\Omega_{\bqp}\rightarrow\Omega$ and 
$g_{\bk,\bq}\rightarrow g=\alpha/\sqrt{2\Omega}$. 
With these simplifications, the vertex correction reduces to 
\begin{equation}\label{eqn:Gamma^(2)(k,q)_Holstein}
    \Gamma^{(2)}(k,q) = \frac{g^{2}T}{N}\sum_{\qp,\sigp}
    D_{0}(\ii\nu_{\mpr}) G_{0}(k-q-\qp) G_{0}(k-\qp). 
\end{equation}
We evaluate the sums directly on finite momentum and frequency grids, thereby 
approximating the vertex in the thermodynamic limit.  
For our calculations, we take $N=24\times24$ and 128 frequencies for a model 
temperature of $T/t = 0.1$. 
The number of Matsubara frequencies was chosen such that the high energy 
cutoff $\hbar\omega_{\cc}\approx 5 W$, where $W$ is the noninteracting bandwidth.

\begin{figure}[t!]
    \includegraphics[width=0.95\columnwidth]{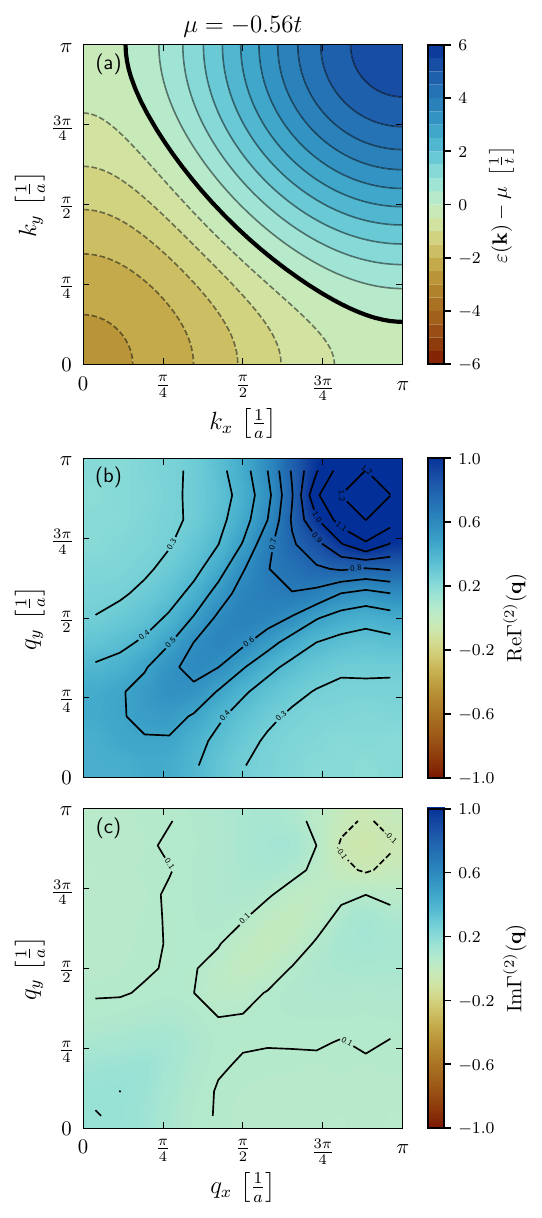}
    \caption{Contour plots of the (a) electronic dispersion $(\varepsilon(\bk)-\mu)/t$ and the (b) real and (c) imaginary parts of the first vertex correction $\Gamma^{(2)}(\bq)$ for $\mu/t=0.56$ and $T/t=0.1$.
    In plot (a), the solid line indicates the Fermi surface contour corresponding to $\varepsilon(\bk)-\mu=0$. 
    In plots (b) and (c), the reported $\Gamma^{(2)}(\bq)$ follows from taking the vertex $\Gamma^{(2)}(\bk,\ii\omega_{n};\bq,\ii\nu_{m})$ that has been evaluated at $\ii\omega_{0}$ and $\ii\nu_{0}$ and then reduced to 
    $\bq$ dependence by carrying out a Fermi surface average over $\bk\in\mrm{FS}$.  
    }
    \label{fig:ek_ReGam_ImGam_mu0p56}
\end{figure}


Figure~\ref{fig:ek_ReGam_ImGam_mu0p56} plots vertex correction as a function of momentum transfer. 
Here, we have simplified the multidimensional vertex function by focusing on the 
lowest Matsubara frequency (i.e., $\ii\omega_{n=0} = \pi T$ and $\ii\nu_{m=0}=0$), 
and performing a Fermi surface average over the fermion wave vectors $\bk$. 
Denoting the simplified vertex correction as $\Gamma^{(2)}({\bf q})$, the    
averaging procedure is given by
\begin{align}
    \Gamma^{(2)}({\bf q})&\equiv  \la \Gamma^{(2)}(\bk,\pi T,\bq,0) \ra_{\bk\in\mrm{FS}} \nonumber
    \\
    &= 
    \frac{\displaystyle
        \sum_{\bk\in\mrm{BZ}} \Gamma^{(2)}(\bk,\pi T,\bq,0)
         \deltaFD(\xi_{\bk})
    }{\displaystyle
        \sum_{\bk\in\mrm{BZ}} 
        \deltaFD(\xi_{\bk})
    } .\label{eq:FS_average}
\end{align}
The wave vectors $\bk$ are restricted to the Fermi surface by use of a ``smeared'' 
delta function $\deltaFD(\xi_{\bk})$ given by 
\begin{equation}
    \deltaFD(x) = -\frac{\D}{\D x}
    \left( \frac{1}{\e^{x/\sigma} + 1} \right)
    = \frac{1}{4\sigma \cosh^{2}\left(\frac{x}{2\sigma}\right)},
\end{equation}
where the broadening parameter $\sigma=\kB T$.
%

Figure~\ref{fig:ek_ReGam_ImGam_mu0p56}(a) shows a contour plot of the underlying 
band structure $\xi_{\bk}$ in the upper quadrant of the first Brillouin zone for 
$\mu/t = -0.56$. 
The thick black line follows the Fermi surface contour $\xi_{\bk}=0$ 
and thin dashed (solid) contour lines are used to plot  $\xi_{\bk}<0$ 
($\xi_{\bk}>0$). 
The values of $t,\,\tp$, and $\mu$ chosen here (to match Ref.~\cite{SchrodiPRB2021}) 
are somewhat typical for modeling a 2D ``cuprate''-like Fermi surface. 
The non-interacting Fermi surface is well nested for transfer vectors near 
$\bq=(\pi,\pi)$, which coincides with the peak in $\chi^{\mrm{CDW}}(\bq)$ seen 
in our HMC results. 
The corresponding $\bq$ dependence of the real and imaginary parts of $\Gamma^{(2)}(\bq)$ 
are displayed in Figs.~\ref{fig:ek_ReGam_ImGam_mu0p56}(b) and 
\ref{fig:ek_ReGam_ImGam_mu0p56}(c), respectively.
(Here, we restrict the plot axes $q_{x},q_{y} \in [0,\pi]$ because the remaining 
quadrants are symmetrically identical.)    
A Gaussian interpolation was used to smooth the $24\times 24$ $\bq$-grid, and 
contours were added to help identify the features and overall magnitude of the 
Fermi surface averaged vertex correction. 
%

It is clear from Fig.~\ref{fig:ek_ReGam_ImGam_mu0p56}(b) that the real part of 
the Fermi-surface averaged vertex correction is of order $O(1)$ near 
$\bq=(\pi,\pi)$. 
For this case, the imaginary part of $\Gamma^{(2)}(\bq)$ 
[Fig.~\ref{fig:ek_ReGam_ImGam_mu0p56}(c)] is relatively small and at most 
$\sim 0.1-0.2$. 
This result implies that an expansion for the vertex $\Gamma({\bf q})
\approx \mathrm{i} g[1 + \Gamma^{(2)}({\bf q}) + \dots]$ involves corrections 
that are on the order of the bare vertex, and thus higher order terms would 
likely be needed to obtain a converged result. 
Consequently, a self-consistent treatment of the first vertex correction in this 
context is likely uncontrolled, and one should assess the strength of the 
second-order diagrams before proceeding. 

\begin{figure*}[t]
    \includegraphics[width=1.0\textwidth]{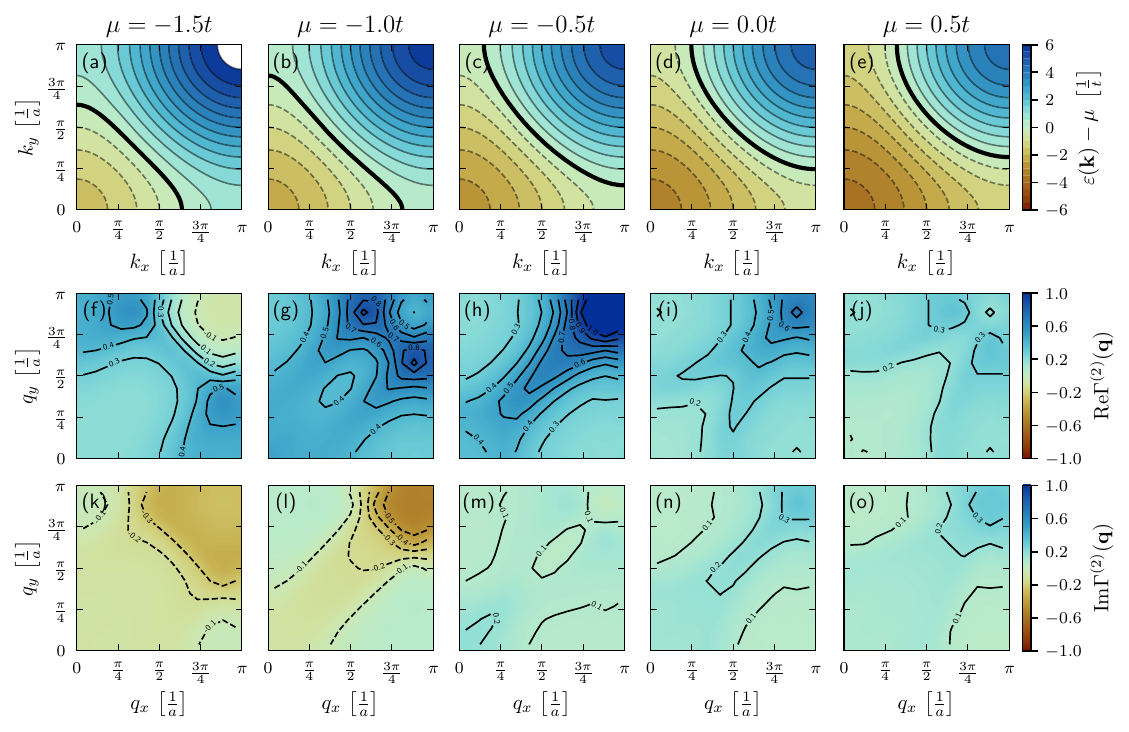}
    \caption{A survey of effects of the Fermi surface features on the first Fermi-surface averaged vertex correction [see Eq.~(\ref{eq:FS_average})]. Results in each column correspond to $\mu/t = -1.5$, $-1$, $-0.5$, 0, and 0.5 reading left to right.  The top row shows contour plots of the bare band dispersion $\xi_\bk$. The thick black line denotes the Fermi surface contour. The thin dashed (solid) lines show contours for $\xi_{\bk} < 0$ ($\xi_{\bk} >0$). The second and third rows show the real and imaginary parts of the Fermi surface averaged vertex $\Gamma^{(2)}(\bq)$ for momentum transfer in the positive quadrant of the first Brillouin zone (FBZ). The values of $\Gamma^{(2)}(\bq)$ at other points in the FBZ can be inferred from $C_4$ symmetry. 
        }\label{fig:ek_ReGam_ImGam}
\end{figure*}    

We now investigate the changes in $\Gamma^{(2)}(\bq)$ as $\mu$ is tuned away 
from $\mu/t = -0.56$ to determine the role of the Fermi surface nesting in the setting the 
magnitude of the correction. 
Figure \ref{fig:ek_ReGam_ImGam} plots  $\xi_{\bk}$ along with  
$\mrm{Re}\Gamma^{(2)}(\bq)$ and $\mrm{Im}\Gamma^{(2)}(\bq)$ by row, but now 
each of the five columns corresponds to a different choice of $\mu/t \in [-1.5, 0.5]$ in steps of $0.5$.
The results for $\mu/t = -0.5$ (middle column) are similar to those shown in 
Fig.~\ref{fig:ek_ReGam_ImGam_mu0p56}; the FS is strongly nested for  
$\bq \approx (\pi,\pi)$ and $\mrm{Re}\Gamma^{(2)}(\pi,\pi)\sim O(1)$.
The nesting condition survives when $\mu/t$ is adjusted by $\pm 0.5$ but shifts 
to different momentum transfers in both cases. 
This fact is evidenced by the strong incommensurate peaks in 
$\mrm{Re}\Gamma^{(2)}(\bq)$ for these values of the chemical potential.  
The peak heights also decrease in these cases but remain large enough to 
invalidate a low-order perturbation expansion in the vertex function. 

The nesting conditions are strongly suppressed for $\mu/t = -1.5$ and $0.5$, as 
shown in the first and fifth columns, respectively. 
In these cases, the band structure begins to resemble a free electron (hole) 
dispersion with a circular Fermi surface. 
(For $\mu/t = -1.5$, the electron-like Fermi surface is more diamond-like.) 
As one might expect, the real and imaginary parts of $\Gamma^{(2)}({\bf q})$ are correspondingly smaller than 1.
For example, $\mathrm{Re}\Gamma^{(2)}({\bf q})$ has a weak peak near ${\bf q} \approx (\tfrac{1}{4},\tfrac{4}{5})\pi$ for the free-electron-like case ($\mu/t = -1.5$), while it has a very weak peak near ${\bf q} \approx (0.9,0.9)\pi$ for the free-hole-like case ($\mu/t = 0.5$). 
We expect that a self-consistent treatment of the first-order vertex corrections may be more controlled in these cases. 
 

%

\subsection{Discussion}

In the previous section, we showed that the first vertex correction acquires a 
momentum anisotropy that follows directly from the geometry of the electronic 
dispersion near the Fermi level. 
This result occurs, even with a bare Holstein coupling and a dispersionless 
Einstein phonon mode, both of which are isotropic in momentum space. 
Improving upon this simplified picture by reintroducing dispersive phonons or 
using the dressed propagators in the vertex diagram could significantly alter 
our conclusion. 
Using dressed propagators $G(k)$ and $D(q)$ instead of $G_{0}(k)$ and $D_{0}(q)$ in 
Eqn.~(\ref{eqn:Gamma^(2)(k,q)_Holstein}) and generating second-order corrections 
to the electron and phonon self-energies [i.e., $\Sigma^{(2)}(k)$ and $\Pi^{(2)}(q)$]
constitute a fully self-consistent evaluation within the vertex-corrected 
Migdal-Eliashberg theory. 
Such a procedure introduces additional $e$-ph-induced renormalization effects on 
\textit{both} the phonons and electrons beyond the usual self-consistent Eliashberg 
formalism.
Even without the vertex correction, treating the electron and phonon 
self-energies on equal footing allows for phonon softening, manifesting as 
a Kohn anomaly in the phonon dispersion at the nesting vector~\cite{DeePRB2019}. 
If such softening were also present in the vertex-corrected theory, the peaks in 
$\Gamma^{(2)}(\bq)$ could be quite strongly affected. 
Our results would reflect the first iteration of such a self-consistent procedure. 

Using HMC simulations as a stand-in for summing all the Feynman diagrams, we 
observed that reintroducing the neglected diagrams greatly favors CDW correlations 
at the expense of superconductivity for the parameters studied.
This result is relatively unsurprising given the sizable dimensionless coupling $\lambda\approx0.88$, which, alongside the frequency of $\Omega/t = 0.4 $, fits into a regime associated with large lattice fluctuations~\cite{Dee2020}.
In the large coupling limit $\lambda \ge 0.5$, these correlations reflect 
bipolaron formation~\cite{EsterlisPRB2018} not captured within the 
framework of Migdal's theory. 
How many additional vertex corrections are needed to describe bipolarons and the 
CDW transition remains unclear.
It is also unclear if a finite number of corrections would be sufficient.

The study of corrections to the electron-phonon vertex has a long history~\cite{Migdal1958,Eliashberg1960,Eliashberg1961,Kostur1994,Benedetti1994,Krishnamurthy1994,Nicol1994,Grimaldi1995,NAsc_I_Pietronero1995,NAsc_II_Grimaldi1995,Freericks1997,Gunnarsson1997,Perali1998,Danylenko2001,Hague2003,Hague2006,EsterlisPRB2018,Chubukov2020,Dee2020}. 
Many of these works studied the ramifications of vertex corrections on 
superconductivity and consider, for instance, how vertex corrections affect the critical temperature and or pairing~\cite{Kostur1994,Benedetti1994,Krishnamurthy1994,Nicol1994,Grimaldi1995,NAsc_I_Pietronero1995,NAsc_II_Grimaldi1995,Freericks1997,Perali1998,Schrodi2020}, the isotope coefficient~\cite{Nicol1994}, 
predictions for non- and antiadiabatic materials like fullerenes~\cite{Nicol1994,Gunnarsson1997}, 
and unconventional superconductivity, often by including them alongside electronic correlations and spin fluctuations~\cite{Grabowski1984,Kostur1994,Krishnamurthy1994,Mierzejewski1998,Zielinski1998,Schrieffer1995,Cappelluti1996,Moskalenko1999,Tsuei2000,Kulic2000}.
This body of literature is extensive, so we have not attempted to review these works comprehensively. 
Instead, we highlight a potential blind spot concerning lurking charge instabilities in vertex-corrected Eliashberg approaches. 
Some findings of the current work have been discussed in the works above to 
varying degrees of rigor.
In particular, when considering problems in 2D, it has been pointed out that Fermi surface 
nesting conditions can invalidate Migdal's approximation, primarily due to geometric singularities appearing in the skeleton diagrams~\cite{Kostur1993}. 
Schrodi \textit{et al}. mention the issue of nesting and several other potential caveats in  Ref.~\cite{Schrodi2020}, which debuts the same state-of-the-art full-bandwidth 
implementation of a vertex-corrected Eliashberg formalism used in 
Ref.~\cite{SchrodiPRB2021}. 
However, as was shown here and by Esterlis \textit{et al}. \cite{EsterlisPRB2018}, comparing Eliashberg-type calculations with nonperturbative methods (e.g., quantum Monte Carlo) can be a vital means of addressing the validity of approximate methods.


\section{Summary \& Conclusion}
Analytic attempts at describing the nature of vertex corrections often entail 
simplifications to the electronic dispersion, typically reducing 
$\varepsilon_{\bk}$ to the single parabolic band of free electrons. 
With modern computing resources, self-consistent solutions of the Migdal-Eliashberg equations, including the \textit{first} vertex correction, are now technically feasible for lattice models with some tight-binding dispersion (e.g., Ref.~\cite{SchrodiPRB2021}). 
However, we have shown that while it may be easier to carry out these 
calculations, the results could be misleading in some instances. 
In particular, we performed HMC simulations using the same 
parameters associated with an unconventional superconducting state in 
Ref.~\cite{SchrodiPRB2021} and instead found a leading charge-density wave 
instability.
We did not attempt to explore the entire parameter space to rule out a 
possible unconventional ground state \textit{somewhere} in the phase diagram. 
However, to our knowledge, the extensive literature on nonperturbative studies of the
Holstein model does not contain any robust evidence for a leading unconventional order parameter thus far. 

To better understand the origin of the $d$-wave order parameter in the 
first-vertex-corrected Eliashberg theory, we evaluated a finite-size approximation 
of the vertex correction.
Due to the relative simplicity of the Holstein model, much of the momentum-space 
structure of $\Gamma^{(2)}(\bq)$ follows directly from the Fermi surface.
For a cuprate-like Fermi surface, we observe peaks in $\Gamma^{(2)}(\bq)$ of 
$O(1)$ near a nesting vector $\bq=(\pi,\pi)$, indicating that higher order 
diagrams may be crucial for these parameters, matching the conclusions 
from our HMC results.

It remains an open question as to when and how vertex corrections should be 
included within Migdal-Eliashberg formalism for specific applications. 
What is clear is that there are scenarios where the perturbative expansion is 
somewhat ill defined and blind to competing phenomena such as bipolaron 
formation. 
The standard Eliashberg formalism below $\Tc$ comes equipped with the usual anomalous 
propagators, which admit a superconducting order parameter but not a competing CDW. 
A possible way forward would be to study the effect of vertex corrections 
in the normal state and tracking both pairing and the CDW correlations 
in the self-consistent Migdal approximation~\cite{MarsiglioPRB1990,DeePRB2019}.
However, to truly understand the physics contained in the corrections, it may be 
necessary to study the diagrams \textit{beyond} the first correction. 
Finding the middle ground where the first correction $\Gamma^{(2)}({\bf q})$ [and 
possibly $\Gamma^{(4)}({\bf q})$] could be used to safely fine-tune predictions is 
a direction of future study.


%




\vspace{0.5cm}
\noindent{\it Acknowledgments} ---  
The authors thank Yan Wang for fruitful discussions about this topic. P.~M.~D. and P.~J.~H. acknowledge support from the U.S. Department of Energy, Office of Science, Office of Basic Energy Sciences, under Award No. DE-FG02-05ER46236. B.C.~S. and S.~J. acknowledge support from the U.S. Department of Energy, Office of Science, Office of Basic Energy Sciences, under Award No. DE-SC0022311.

\bibliography{references}
\end{document}